\documentclass[aps,pre,twocolumn,showpacs,groupedaddress]{revtex4}

\usepackage{graphicx}
\usepackage{amssymb}

\newcommand{\Eq}[1]{Eq.~(\ref{#1})}
\newcommand{\eq}[1]{(\ref{#1})}

\newcommand{\Eqs}[1]{Eqs.~(\ref{#1})}

\newcommand{\Fig}[1]{Fig.~\ref{#1}}

\newcommand{\Sec}[1]{Section~\ref{#1}}

\newcommand{\ie}{{\em i.e.~}}

\newcommand{\dg}{\dot{\gamma}}

\newcommand{\rr}{\mathbf{r}}

\newcommand{\mm}{\mathrm{m}}
\newcommand{\kk}{\mathrm{K}}
\newcommand{\cc}{\mathrm{c}}
\newcommand{\pp}{\mathrm{p}}
\newcommand{\kt}{k_{\mathrm{B}}T}

\newcommand{\rot}{\mbox{$\hat{\mathcal{R}}$}}

\begin{document}

\title{Multi-particle collision dynamics modeling of
       viscoelastic fluids 
      }

\author{Yu-Guo Tao
  \footnote{Present address: Department of Chemistry, 
            University of Toronto, Toronto, ON, M5S 3H6, Canada}
       }
  \email[]{ytao@chem.utoronto.ca}
\author{Ingo O. G\"otze}
  \email[]{i.goetze@fz-juelich.de}
\author{Gerhard Gompper}
  \email[]{g.gompper@fz-juelich.de}
 \affiliation{Theoretical Soft Matter and Biophysics Group,
             Institut f\"ur Festk\"orperforschung,
             Forschungszentrum~J\"ulich,
             D-52425~J\"ulich,~Germany}
\date{\today}


\begin{abstract}

In order to investigate the rheological properties of viscoelastic fluids
by mesoscopic hydrodynamics methods,
we develop a multi-particle collision dynamics (MPC) model for a
fluid of harmonic dumbbells. The algorithm consists of alternating
streaming and collision steps. The advantage of the harmonic interactions
is that the integration of the equations of motion in the streaming
step can be performed analytically. Therefore, the algorithm is computationally
as efficient as the original MPC algorithm for Newtonian fluids.
The collision step is the same as in the original MPC method.
All particles are confined between two solid walls moving oppositely, 
so that both steady and oscillatory shear flows can be investigated.
Attractive wall potentials are applied to obtain a nearly uniform density 
everywhere in the simulation box.
We find that both in steady and oscillatory shear flow, a boundary layer
develops near the wall, with a higher velocity gradient than in the 
bulk. The thickness of this layer is proportional to the average
dumbbell size. We determine the zero-shear
viscosities as a function of the spring 
constant of the dumbbells and the mean free path. For very high shear
rates, a very weak ``shear thickening'' behavior is observed. 
Moreover, storage and loss moduli are calculated in oscillatory shear, 
which show that the viscoelastic properties at low and moderate frequencies 
are consistent with a Maxwell fluid behavior. We compare our results with
a kinetic theory of dumbbells in solution, and generally find good
agreement. 

\end{abstract}

\pacs{47.11.-j,    
      83.60.Bc,    
      66.20.+d}    

\maketitle


\section{Introduction}

It is the characteristic feature of soft matter systems that a macromolecular
component of nano- to micrometer size is dispersed in a solvent of much
smaller molecules. The mesoscopic length scale of the dispersed
component implies that crystalline phases have a very small shear modulus -- 
which roughly scales
like the inverse of the third power of the structural length scale --
and that both crystalline and fluid phases are characterized by long 
structural relaxation times. Soft matter systems have
therefore interesting dynamical properties, because the time scale of
an external perturbation can easily become comparable with the
intrinsic relaxation time of the dispersed macromolecules.

One of the unique properties of soft matter is its viscoelastic
behavior\cite{lars99b}. Due to the long structural relaxation time, 
the internal
degrees of freedom cannot relax sufficiently fast in an oscillatory
shear flow, so that there is some elastic restoring force which
pushes the system back to its previous state. A very well studied
example of viscoelastic fluids are polymer solutions and polymer
melts\cite{ferr80b,doi86,lars99b}. In the case of polymer melts,
the characteristic time scale 
is given by the reptation time, {\em i.e.} by the time
it takes a chain to slide by its contour length along the tube formed 
by other polymer chains\cite{doi86}. 

In order to bridge the length- and time-scale gap between the solvent 
and macromolecular or colloidal scales, several mesoscopic simulation
techniques -- such as the lattice-Boltzmann method, dissipative-particle 
dynamics (DPD), and multi-particle collision dynamics (MPC) -- have 
been suggested in recent 
years, and are in the process of being developed further.  The idea of 
all these methods is to strongly simplify the microscopic dynamics in 
order to gain computational efficiency, but at the same time to exactly 
satisfy the conservation laws of mass, momentum and energy, so that 
hydrodynamic behavior emerges naturally on larger length scales. 

We will focus here on the multi-particle collision dynamics (MPC) 
technique\cite{male99,male00a,lamura2001}, also called stochastic rotation 
dynamics\cite{ihle01} (SRD), originally developed for Newtonian fluids. 
This particle-based hydrodynamics method 
consists of alternating streaming and collision steps. In the streaming
step, point particles move ballistically. In the collision step, particles are
sorted into the cells of a simple cubic (or square) lattice. All particles
in a cell collide by a rotation of their velocities relative to the 
center-of-mass velocity around a random axis\cite{male99}. A random 
shift of the cell lattice is performed before each collision step in
order to restore Galilean invariance\cite{ihle01}. 
This method has been applied very successfully to study the hydrodynamic 
behavior of many complex fluids, such as polymer 
solutions in equilibrium\cite{gg:gomp05h,lee06} and 
flow\cite{webster2005,ryde06,gg:gomp06d}, 
colloidal dispersions\cite{padd04,hech05}, 
vesicle suspensions\cite{gg:gomp04h,gg:gomp05c}, 
and reactive fluids\cite{tucc04,eche07}.

The viscoelastic behavior of polymer solutions leads to many unusual
flow phenomena, such as shear-induced phase 
separation\cite{helf89,onuk89,miln93}, 
viscoelastic phase separation\cite{tana00b}, 
and elastic turbulence\cite{groi00}. 
A coarse-grained description of viscoelastic fluids is necessary 
in order to obtain a detailed understanding of the role of elastic forces 
in such flow instabilities. 

However, there is a second level of complexity in soft matter system, 
in which a colloidal component is dispersed in a solvent, which is 
itself a complex fluid. Examples are spherical or rod-like colloids 
dispersed in polymer solutions or melts, which are exposed to a shear 
flow\cite{lyon01,suen02,hwan04,scir04,verm05}. Shear flow can induce
particle aggregation and alignment in these systems. This is important,
for example, in the processing of nanocomposites\cite{verm05}. 

The aim of this paper is therefore the development of a MPC algorithm,
which is able to describe viscoelastic phenomena, but at the same time
retains the computational simplicity of standard MPC for Newtonian
fluids, and thereby allows to take advantage of this mesoscale simulation
for the investigation of flow instabilities as well as suspensions with
viscoelastic solvents. We show that this goal can be achieved by 
replacing the point particles of standard MPC by harmonic dumbbells.
In order to obtain a strong elastic contribution, we consider a fluid,
which consists of dumbbells only. However, it is of course straightforward
to mix dumbbells with a point-particle solvent. A similar idea has been
suggested recently for DPD fluids\cite{somf06}.


\section{The Model}
\label{sec:model}

\subsection{Algorithm}

In our MPC model, we consider $N_\pp$ point particles of mass $\mm$, 
which are pairwise connected by a harmonic potential 
$\mathbf{V}(\rr_{1},\rr_{2})=\frac{1}{2} \mathrm{K} (\mathbf{r}_{1} - \mathbf{r}_{2} )^2$
to form dumbbells, where $\mathrm{K}$ is the spring constant.
The center-of-mass position $\mathbf{r}^{\cc}_{i}$ and velocity 
$\mathbf{v}^{\cc}_{i}$
for each dumbbell $i$, with $i=1,2,...,N_\pp/2$, are represented by
\begin{equation}
 \mathbf{r}^{\cc}_{i} = \frac{1}{2} (\mathbf{r}_{i1} + \mathbf{r}_{i2} ) \;; \; \; \;
 \mathbf{v}^{\cc}_{i} = \frac{1}{2} (\mathbf{v}_{i1} + \mathbf{v}_{i2} ) \;.
\end{equation}
Here $\mathbf{r}_{i1}$, $\mathbf{r}_{i2}$ and $\mathbf{v}_{i1}$, 
$\mathbf{v}_{i2}$ denote the position and velocity of the two point 
particles composing a dumbbell $i$, respectively.

The MPC algorithm consists of two steps, streaming and 
collisions\cite{male99,male00a,malevanets2004}.
In the streaming step, within a time interval $h$, 
the motion of all dumbbells is governed by Newton's equations of motion,
\begin{equation}
  \mathrm{m}^\cc \frac{d\mathbf{v}^{\cc}_{i}}{dt} = \mathbf{f}^{\cc}_{i} \;; \; \;
  \frac{d\mathbf{r}^{\cc}_{i}}{dt} = \mathbf{v}^{\cc}_{i} \;,
\end{equation}
where $\mathrm{m}^\cc=2\mm$ is the mass of a dumbbell,
and $\mathbf{f}^{\cc}_{i}$ is the total external force on dumbbell $i$.
We consider only constant force fields.
The center-of-mass positions and velocities of dumbbells are then given 
by a simple ballistic motion.
The evolution of the relative coordinates of each dumbbell 
are determined by the harmonic interaction potential, so that
\begin{eqnarray}
 \label{rr}
 \mathbf{r}_{i1}(t+h) - \mathbf{r}_{i2} (t+h)
 & = & \mathbf{A}_{i}(t) \cos(\omega_0 h) \nonumber \\ && + \mathbf{B}_{i}(t) \sin(\omega_0 h)\;; \\
 \label{rv}
 \mathbf{v}_{i1}(t+h) - \mathbf{v}_{i2} (t+h) 
 & = & - \omega_0 \mathbf{A}_{i}(t) \sin(\omega_0 h)  \nonumber \\ 
 & & + \omega_0 \mathbf{B}_{i}(t) \cos(\omega_0 h) \;,
\end{eqnarray}
with angular frequency $\omega_0 = \sqrt{2\mathrm{K} / \mathrm{m}}$.
The vectors $\mathbf{A}_{i}$ and $\mathbf{B}_{i}$ are 
different for each time step, and are 
calculated from the relative positions and velocities 
of the point particles of dumbbell $i$ before the streaming step,
\begin{equation}
  \mathbf{A}_{i}(t) = \mathbf{r}_{i1}(t) - \mathbf{r}_{i2} (t)\;; \; \;
  \mathbf{B}_{i}(t) = \frac{1}{\omega_0}\left(\mathbf{v}_{i1}(t) - \mathbf{v}_{i2} (t)\right) \;.
\end{equation}
In the MPC algorithm described here, $\mathbf{r}^{\cc}$, 
$\mathbf{v}^{\cc}$, $\mathbf{A}$ and $\mathbf{B}$ are continuous variables, 
evolving in discrete increments of time.
In the absence of shear flow, the average length of the dumbbell is
$\mathrm{r_0}^{(d)} \equiv \sqrt{\langle \mathrm{r}^2 \rangle_{\rm eq}}=
\sqrt{d~\kt / \mathrm{K}}$ for a $d$-dimensional system.

In the collision step, the point particles are sorted into the cells of 
a cubic lattice with lattice constant $a_0$.
Multi-particle collisions are performed for all particles in a cell $J$,
by the same SRD algorithm\cite{male99} as for point particle fluids. The 
velocity of each particle relative 
to the center-of-mass velocity $\mathbf{v}_{\mathrm{cm},J}$ of the cell
is rotated around a randomly chosen axis by a fixed angle $\alpha$, 
\begin{equation}
\label{vcm}
 \mathbf{v}'_j(t+h) = \mathbf{v}_{\mathrm{cm},J} + \rot(\alpha) \left[\mathbf{v}_j(t+h)-
  \mathbf{v}_{\mathrm{cm},J} \right]\;,
\end{equation}
where $\rot(\alpha)$ is a stochastic rotation matrix, and 
\begin{equation}
\label{vcm2}
 \mathbf{v}_{\mathrm{cm},J} =  \sum^{N_J}_{j=1}\mathbf{v}_j/N_J\;,
\end{equation}
with $N_J$ the number of particles within cell $J$.
This step guarantees that each particle changes the direction as well as 
the magnitude of its velocity during the multi-particle collisions, while 
the local momentum and the kinetic energy are conserved. 
Random shifts are applied in each direction, so that the Galilean 
invariance is ensured 
even in case of small mean free path\cite{ihle01,ihle2003}. 

In order to describe Couette or oscillatory shear flow, the system is 
confined within two parallel hard walls in the $y$ direction, 
which are moving oppositely along the $x$ direction. 
Here, $L_x$, $L_y$ and $L_z$ are used to denote the dimension of the 
simulation box along the corresponding directions.
For a steady shear flow, the shear rate is given by $\dg =2 v_{\mathrm{wall},x} / L_y$,
with $v_{\mathrm{wall},x}$ the $x$ component of 
the velocity of the wall moving along the positive direction.
Periodic boundary conditions are applied in the $x$ and $z$ direction, 
bounce-back boundary condition in the $y$ direction.
The system is therefore divided into $L_x/a_0$ and $L_z/a_0$ cells 
in the $x$ and $z$ directions (parallel to the walls),
but $L_y/a_0 +1$ cells in the $y$ direction because of the random shifts. 
At the walls, for collision cells which are not completely filled by particles, 
extra virtual point particles are added to conserve the monomer number 
density, $\rho$, defined by the average number of monomers per 
cell\cite{lamura2001}. 
In principle, the velocities of the virtual particles can be drawn from 
a Maxwell-Boltzmann distribution of average velocity equal to the wall 
velocity and variance $\sqrt{\kt / \mathrm{m}}$, where $\kt$ the bulk 
temperature.  In the simulation code, it is not necessary to sample the 
velocity of virtual wall particles individually.
A random vector from Maxwell-Boltzmann distribution with wall velocity 
and variance $\sqrt{(\rho - n)\kt / \mathrm{m}}$ is then used instead 
of the contribution of the entire virtual particles in the cell, 
where $n$ is the number of real particles in that cell.
For point particles, the combination of bounce-back boundary condition and 
virtual wall particles has been shown to guarantee no-slip boundary 
condition to a very good approximation\cite{lamura2001}.

\subsection{Thermostats}

In order to keep the system temperature constant, various thermostats 
can be employed.  In the first case, the MPC method with collisions by 
stochastic rotations (MPC-SRD)
of relative velocities is augmented by velocity rescaling. The simulation 
box is subdivided into $L_y/a_0$ layers parallel to the walls.
In each layer, the new velocity $\mathbf{v}'_j$ of each particle 
$j$ in cell $J$ is obtained by rescaling the velocity relative to the 
center-of-mass velocity of that cell,
\begin{equation}
\label{th1}
 \mathbf{v}'_j = \mathbf{v}_{\mathrm{cm},J} +
 ( \mathbf{v}_j - \mathbf{v}_{\mathrm{cm},J} ) \sqrt{\frac{k_{\mathrm{B}}T}{k_{\mathrm{B}}T'}}\;.
\end{equation}
Here $k_{\mathrm{B}}T'$ is calculated from the actual velocity distribution 
\begin{equation}
\sum_{J\in {\rm layer}} \sum^{N_J}_{j=1} \frac{1}{2} \mathrm{m} ( \mathbf{v}_j - \mathbf{v}_{\mathrm{cm},J} )^2 =
 ( \sum_{J\in {\rm layer}} N_J - \tilde{N}_{\rm layer} ) k_{\mathrm{B}}T' \;,
\end{equation}
where $N_J$ denotes the number of particles in cell $J$ and 
$\tilde{N}_{\rm layer}$ the number of cells which contains particles 
within a layer. 

In the second case, the Anderson's thermostat version of MPC, denoted MPC-AT, 
is applied\cite{noguchi2007,gg:gomp02c}.
This thermostat employs a different collision rule instead of Eq.~(\ref{vcm}).
In the MPC-AT$-a$ version of the algorithm (without angular momentum 
conservation, compare Sec.~\ref{sec:angular} below),
the new velocities of point particles in the collision step are assigned 
as\cite{noguchi2007}
\begin{equation}
\mathbf{v}'_j = \mathbf{v}_{\mathrm{cm},J} + \mathbf{v}^{\mathrm{ran}}_j 
- \sum^{N_K}_{k=1}\frac{\mathbf{v}^{\mathrm{ran}}_k}{N_K} \;.
\end{equation}
Here $\mathbf{v}^{\mathrm{ran}}_j$ is a velocity chosen from the 
Maxwell-Boltzmann distribution and $N_K$ the number of particles within 
cell $\kk$.  Instead of energy conservation in MPC, the temperature is kept 
constant in MPC-AT.

\subsection{Angular Momentum Conservation}
\label{sec:angular}

The standard MPC algorithm as well as the Anderson thermostat version 
do not conserve angular momentum.
It has been shown recently\cite{gg:gomp07h} that this lack of 
angular-momentum conservation may lead to quantitative or even 
qualitative incorrect results, like non-physical torques in circular 
Couette flows.
We therefore also consider the angular-momentum conserving modification of 
MPC-AT\cite{noguchi2007,gg:gomp07h}, denoted MPC-AT$+a$.
Here, the velocities in the collision step are calculated by
\begin{eqnarray}
\mathbf{v}'_j & = & \mathbf{v}_{\mathrm{cm},J} + \mathbf{v}^{\mathrm{ran}}_j 
- \sum^{N_K}_{k=1}\frac{\mathbf{v}^{\mathrm{ran}}_k}{N_K} \\
& + & \left\{ \mathrm{m}~\mathbf{\Pi}^{-1}~ \sum^{N_K}_{k=1} 
\left(\mathbf{r}_k - \mathbf{r}_{\mathrm{cm},K}\right) 
\times \left( \mathbf{v}_k - \mathbf{v}^{\mathrm{ran}}_k \right) \right\}
\nonumber \\
& \times & \left(\mathbf{r}_j - \mathbf{r}_{\mathrm{cm},J}\right) \;,  \nonumber
\end{eqnarray}
where $\mathbf{\Pi}$ and $\mathbf{r}_{\mathrm{cm},J}$ denote the 
moment-of-inertia 
tensor and the center of mass of particles in the cell, respectively.

\subsection{Wall Potential}

\begin{figure}[ht]
  \begin{center}
    \includegraphics[clip]{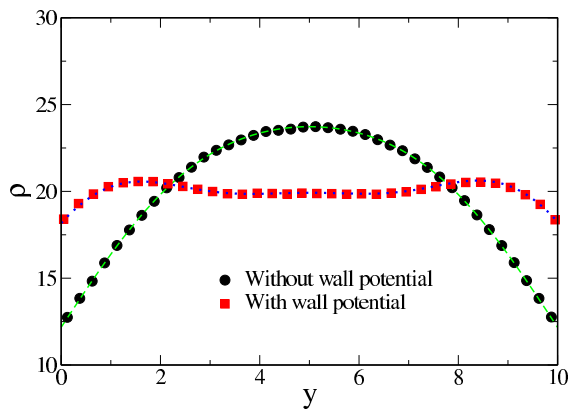}
  \end{center}
  \caption{
    \label{den}
     (Color online) 
     Monomer density profiles with (squares) and without (circles) 
     attractive wall potentials applied along the wall direction 
     when particles approach close to walls.
     The dashed and dotted lines are the theoretical prediction 
     described in \Eq{wall} and \Eq{wall2}, respectively.
     The spring constant of dumbbells and the collision time are 
     $\mathrm{K}=0.2$ and $h=0.02$, respectively. 
     Both simulations are with absence of shear flow.  
  }
\end{figure}

In the absence of shear flow, the monomer density profile $\rho(y)$ can be 
calculated from the interaction potentials $V$ of the dumbbells,
\begin{eqnarray}
\label{wall}
 \rho(y) & = & \frac{1}{Z} \int^{L_y}_{0} d y'~ 
 \mathrm{e}^{-\frac{1}{2} \frac{\mathrm{K}}{\kt} (y-y')^2} \;,\\
 \label{wall0}
 \rho(y) / \rho_\mathrm{b} & = & \frac{1}{2} 
     \left[ \mathrm{erf}\left(\sqrt{\kk /2 \kt}~y\right) \right. \nonumber \\
     & + & \left. \mathrm{erf}\left(\sqrt{\kk /2 \kt}~(L_y-y)\right)\right] \;,
\end{eqnarray}
where $\rho_\mathrm{b}$ is the bulk monomer density, $Z$ the partition function,
and \textit{erf} the error function.
\Fig{den} shows excellent agreement of the theoretical prediction 
\eq{wall0} with simulation data.
The particles are not equally distributed along the wall direction;
instead, at both walls, the density is only half of the bulk density. 
In order to reduce possible slip effects, 
it seems desirable to make the particle distribution as uniform as possible.
An attractive potential is therefore applied 
when the center-of-mass position of the dumbbells approaches one of the walls, 
\begin{eqnarray}
\label{wall1}
 V_{\mathrm{wall}}(y_{i1},y_{i2}) & = & -2c_2~\kt~
 \left(1 - \frac{y_{i1}+y_{i2}}{2c_1\mathrm{r}_0^{(1)}}~\right)  \nonumber \\
 && \mathrm{for}~~\frac{y_{i1}+y_{i2}}{2} \leq c_1\mathrm{r}_0^{(1)} \;; \nonumber \\
 V_{\mathrm{wall}}(y_{i1},y_{i2}) & = & -2c_2~\kt~
 \left(1 - \frac{2L_y - y_{i1} - y_{i2}}{2c_1\mathrm{r}_0^{(1)}}~\right)  \nonumber \\ 
 && \mathrm{for}~~\frac{y_{i1}+y_{i2}}{2}  \geq L_y- c_1\mathrm{r}_0^{(1)} \;,
\end{eqnarray}
where $\mathrm{r}_0^{(1)}=\sqrt{\kt / \kk}$ is the one-dimensional average 
extension of a dumbbell.  The density profile is now given by
\begin{equation}
\label{wall2}
 \rho(y) = \frac{1}{Z} \int^{L_y}_{0} d y'~ 
 \mathrm{e}^{-\frac{1}{2} \frac{\mathrm{K}}{\kt} (y-y')^2} 
 \mathrm{e}^{- V_{\mathrm{wall}}(y,y') }  \;.
\end{equation}
The advantages of the piecewise linear form \eq{wall1} of the wall 
potential are twofold.  Firstly and most importantly, it allows for 
an analytical integration of the equations of motion during the 
streaming step.  Secondly, the density profile in the absence of 
flow can again be calculated analytically (see Appendix for details).

\begin{figure}[ht]
  \begin{center}
    \includegraphics[clip]{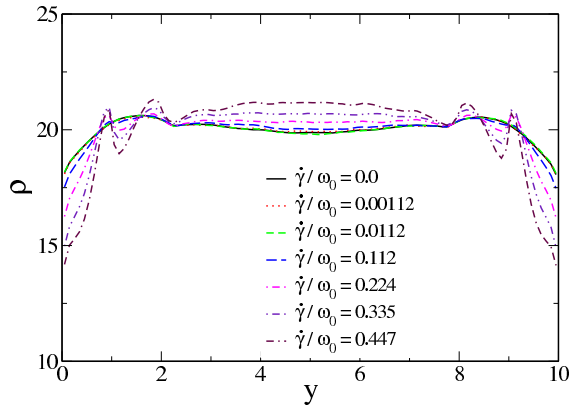}
  \end{center}
  \caption{
    \label{rho}
     (Color online) 
     Monomer density profiles at various dimensionless shear rates 
     $\dg / \omega_0$, ranging from $\dg / \omega_0=0.0$ to $0.447$, 
     when attractive wall potentials are applied.
     The spring constant of dumbbells and the collision time are
     as same as those in \Fig{den}. The small ripples in the profile at
     large $\dot\gamma/\omega_0$ are due to inhomogeneities in the
     temperature profile, since velocity rescaling is not sufficiently
     efficient at high shear rates. The ripples to not appear for MPC-AT.
  }
\end{figure}

The simulated density profile shows excellent agreement with
the analytical solution of \Eqs{wall1} and \eq{wall2} (see Appendix).
The factors $c_1$ and $c_2$ are chosen to obtain a nearly uniform density 
distribution.  This is achieved for $c_1=1.3$ and $c_2=0.4$.
As shown in \Fig{den}, the densities of point particles at both wall 
boundaries deviate by less than $10\%$ lower from the bulk value, when 
the attractive wall potentials are applied.
Simulations are also performed on systems of dumbbells with various 
spring constants, ranging from $\kk a_0^2/\kt = 0.1$ to 
$\kk a_0^2/\kt = 5.0$, in the absence of shear flow.
It is found that for the given values of $c_1$ and $c_2$, the density 
profiles are essentially independent of the spring constant of the 
dumbbell in the range $0.1 < \kk a_0^2/\kt < 1.0$.
In \Fig{rho}, we plot density profiles in shear flow.
At lower shear rates, \ie $\dg /\omega_0 \leq 0.1$, nearly identical 
profiles are obtained as without flow.
For higher shear rates, deviations of the density profile from
the equilibrium profile become significant. 
Nevertheless, these profiles are still more uniform than those without an 
attractive wall potential.  
Our investigations are mainly 
focusing on relatively low shear rates, where
the non-uniformity of the density profile is not significant.

\subsection{Stress Tensor and Shear Viscosity}
\label{sec:stress}

In the MPC model, the viscosity $\eta$ consists of a 
kinetic and collisional contribution\cite{kikuchi2003,tuez03}.
At steady shear rates, 
with flow along the $x$ direction and gradient along the $y$ direction, 
$\eta$ is calculated by measuring the $xy$ component of the stress tensor, 
$\sigma_{xy} = \sigma^{\mathrm{kin}}_{xy} + \sigma^{\mathrm{col}}_{xy}$, 
so that $\eta=\sigma_{xy} / \dg$.

In the streaming step, $\sigma^{\mathrm{kin}}_{xy}$ is proportional to 
the flux of the $x$ momentum crossing a plane normal to the $y$ direction.
Since the stress tensor is independent of the position of the plane, 
we choose $y=0$ or $y=L_y$ to measure the momentum transfer. 
In two-dimensional simulations, 
\begin{equation}
\label{stresskin}
 \sigma^{\mathrm{kin}}_{xy} = \frac{\mathrm{m}}{L_x h} 
 \sum^{N_1}_{i=1} \left[ v'_{x,i}(t_w) - v_{x,i}(t_w) \right]\;,
\end{equation}
where $t_w \in [t,t+h]$ is the time at which particle $i$ bounces back
from the wall, $v_{x,i}(t_w)$ and $v'_{x,i}(t_w)$ are the velocities just
before and after the collision with the wall, and $N_1$ denotes the 
number of particles which hit one of the walls in the time interval
$[t,t+h]$. In the collision step, 
particles close to the wall will change their velocities 
due to the multi-particle collisions with virtual wall particles 
with average velocity $v_x=\pm \frac{1}{2}\dg L_y$,
so that
\begin{equation}
\label{stresscol}
 \sigma^{\mathrm{col}}_{xy} = \frac{\mathrm{m}}{L_x h} 
    \sum^{N_2}_{i=1} \left[ v'_{x,i}(t+h) - v_{x,i}(t+h) \right]\;.
\end{equation}
Here $N_2$ denotes the number of particles 
which have multi-particle collisions with virtual particles, 
while $v_{x,i}(t+h)$ and $v'_{x,i}(t+h)$ are the velocities of particle $i$ 
before and after the collision step, respectively.
In our simulations, $N_2$ is found to be much larger than $N_1$ 
for small collision times $h$, 
indicating that the collisional part dominates the shear viscosity.
Simulations are first performed on a system of pure point-like fluid 
particles to verify the measurement of the zero-shear viscosity 
from \Eqs{stresskin} and \eq{stresscol}.  We get perfect agreement with 
the theoretical predictions\cite{kikuchi2003,tuez03,ripoll2005} for $\eta$.

The shear viscosity can also be measured from system 
under Poiseuille flow\cite{lamura2001,ripoll2006} by
\begin{equation}
\label{Poiseuille}
 \eta = \frac{\rho g L_y^2}{8v_\mathrm{max}}\;,
\end{equation}
where $g$ is the gravitation field, and $v_\mathrm{max}$ the maximum 
flow velocity.

\subsection{Storage and Loss Moduli}

In an oscillatory shear flow, the shear rate $\dg(t)$ is time-dependent,
\begin{equation}
\label{dgt}
 \dg(t) = \gamma_0~\omega~\cos{(\omega t)}\;,
\end{equation}
where $\gamma_0$ and $\omega$ are 
the strain amplitude and the oscillation frequency, respectively.
Note that the frequency $\omega$ in \Eq{dgt} is independent of 
the angular frequency $\omega_0$ of harmonic dumbbells in \Sec{sec:model}.
In our simulations, we choose $\gamma_0 \ll 1$ in order to investigate 
the linear viscoelastic regime.
The stress tensor is divided into two contributions, 
the viscous part $\sigma'$ and the elastic part $\sigma''$, so 
that\cite{lars99b,macosko_book} 
\begin{eqnarray}
\label{ss}
 \sigma_{xy}(t) & = & \sigma' \sin{(\omega t)} + \sigma'' \cos{(\omega t)} \nonumber \\
           & = & \gamma_0 \left[G'(\omega) \sin{(\omega t)} + G''(\omega) \cos{(\omega t)}\right]\;,
\end{eqnarray}
where $G'$ is the storage modulus, which measures the in-phase storage 
of the elastic energy, and $G''$ is the loss modulus, which measures 
the out-of-phase energy dissipation.
For a simple Maxwell fluid, $G'$ and $G''$ are given by\cite{macosko_book}
\begin{eqnarray}
\label{mg1}
 G' & = & G^*
 \frac{(\omega/\omega^*)^2}{1+(\omega/\omega^*)^2} \;;  \\
\label{mg2}
 G'' & = & G^*
 \frac{\omega/\omega^*}{1+(\omega/\omega^*)^2} \; \;,
\end{eqnarray}
where $\omega^*$ is a characteristic relaxation frequency, and $G^*$ 
is a characteristic shear modulus. 
In the limit of $\omega \ll \omega^*$, the loss modulus is 
$G''=\eta~\omega$, where $\eta$ is the zero-shear viscosity.

\subsection{Kinetic Theory of Dumbbells in Solution}

In order to estimate the rheological properties of our model fluid, we 
modify the kinetic theory for dilute solutions of elastic 
dumbbells \cite{bird87}.
For Hookean dumbbells in a solvent, the viscosity $\eta_0$, the storage 
modulus $G_0^\prime$ and the loss modulus $G_0^{\prime\prime}$ are given 
by\cite{bird87}
\begin{equation}
\eta_0 = \eta_s + \frac{\rho}{2} \, \frac{k_BT}{\omega_s},
\label{visc_theo0}
\end{equation}
\begin{equation}
G_0^\prime=\frac{\rho k_BT}{2} \, \frac{(\omega / \omega_s)^2}{1+(\omega / \omega_s)^2},
\end{equation}
\begin{equation}
G_0^{\prime\prime}=\eta_s\omega+\frac{\rho k_BT}{2} \, 
                 \frac{\omega / \omega_s}{1+(\omega / \omega_s)^2},
\end{equation}
where
\begin{equation}
\omega_s=\frac{4\kk}{\zeta_s}
\end{equation}
with solvent viscosity $\eta_s$ and friction coefficient $\zeta_s$ of a 
monomer. 
Moreover, the expectation value for the square of the monomer separation, 
divided by its equilibrium value, is given by\cite{bird87}
\begin{equation}
\frac{\langle r^2 \rangle}{\langle r^2 \rangle_{\rm eq}}=1+\frac{2}{3}(\dot{\gamma} / \omega_s)^2.
\end{equation}

In Ref.~\onlinecite{bird87}, the friction coefficient is obtained from 
Stokes' law for a bead of radius $r$ in the solvent, \ie 
$\zeta_s=6 \pi \eta_s r$.
However, in the MPC dumbbell fluid, there exists no explicit solvent and 
the monomers are point particles instead of spheres.
Nevertheless, the motion of the monomers is governed by the friction 
caused by the surrounding monomers which can be considered to take the 
role of the solvent.
Using $\zeta=k_BT/D$, which follows from the Stokes-Einstein relation, 
we can thus relate the friction to the diffusion constant $D$ of a 
MPC fluid of point particles with the same monomer density.
Similarly, we substitute the viscosity of the solvent, $\eta_s$, by the 
corresponding viscosity $\eta_{\rm MPC}$ of a MPC fluid of point 
particles. 
Theoretical expression for $\eta_{\rm MPC}$ and $D$ for the different
collision methods can be found in
Refs.~\onlinecite{kikuchi2003,tuez03,gg:gomp07h,gg:gomp07xxe} and
Refs.~\onlinecite{ripoll2005,tuez06,gg:gomp07xxe}, respectively.
The zero-shear viscosity then reads
\begin{equation}
\eta = \eta_{\rm MPC} + \frac{\rho}{2} \, \frac{k_BT}{\omega_H},
\label{eq:visc_theo_MPC}
\end{equation}
where we have introduced
\begin{equation}
\label{eq:omega_H}
\omega_H=\frac{4\kk}{\zeta}=\frac{4D\kk}{k_BT}.
\end{equation}
Note that the limit $\kk \rightarrow \infty$ corresponds to a MPC fluid of 
$N_p/2$ point particles of mass $\mathrm{m}^\cc$. Here, the second term in 
Eq.~(\ref{eq:visc_theo_MPC}) vanishes, and since 
$\eta_{\rm MPC}(\rho/2,2\mathrm{m}) \approx \eta_{\rm MPC}(\rho,\mathrm{m})$ for not too 
small $\rho$ and sufficiently small $h$ (so that the collisional part of 
the viscosity dominates), 
the viscosity resulting from this simple theory approaches the correct 
value in this limit.

Consequently, we use the same substitutions for the storage and loss 
modulus, and for the average dumbbell extension, and obtain
\begin{equation}
G^\prime=\frac{\rho k_BT}{2} \, \frac{(\omega / \omega_H)^2}{1+(\omega / \omega_H)^2}
\end{equation}
\begin{equation}
G^{\prime\prime}=\eta_{\rm MPC}\omega 
    + \frac{\rho k_BT}{2} \, \frac{\omega / \omega_H}{1+(\omega / \omega_H)^2},
\end{equation}
and
\begin{equation}
\frac{\langle r^2 \rangle}{\langle r^2 \rangle_{\rm eq}}=1+\frac{2}{3}(\dot{\gamma} / \omega_H)^2.
\label{eq:R2}
\end{equation}
We emphasize that the above expressions only serve as a semi-quantitative
description of the MPC dumbbell fluid. For example, the employed expressions
for the diffusion constant neglect hydrodynamic interactions, which
become important for small time steps $h$.


\section{Results}
\label{sec:results}

\subsection{Dimensionless Variables and Parameters}

In the remainder of this article, we introduce dimensionless 
quantities by measuring 
length in unit of the lattice constant $a_0$, 
mass in unit of the dumbbell mass $\mm^\cc$,
time in units of $a_0\sqrt{\mm^\cc/\kt}$,
velocity in units of $\sqrt{\kt/\mm^\cc}$,
monomer number density $\rho$ in units of $a_0^{-d}$, where $d$ is the 
spatial dimension, and the spring constant $\kk$ in units of $\kt/a^2_0$.
The shear rate $\dg$ and all kinds of frequencies  
are measured in units of $\sqrt{\kt/\mm^\cc a^2_0}$.
Finally the viscosity $\eta$ is in units of $\sqrt{\mm^\cc \kt/a^2_0}$,
and the storage modulus $G'$ and the loss modulus $G''$ are in units of
$\kt/a^3_0$.
In these dimensionless units, the mean free path $\lambda$ (in units
of the lattice constant) becomes equivalent to the time step $h$.

In our simulations, harmonic dumbbells with $N_\mathrm{p}$ point particles 
are initially placed in a two- or three-dimensional rectangular box at random. 
We choose the average number density of point particles $\rho=20$ 
and $L_x=50$ for all two-dimensional simulations
which results in $N_\mathrm{p}=1000L_y$.
The collision time ranges from $h=0.01$ to $h=0.2$, 
while the spring constant ranges from $\kk=0.1$ to $\kk=5.0$.
The rotational angle is chosen $\alpha=90^\mathrm{o}$ 
and $\alpha=130^\mathrm{o}$ for two- or three-dimensional simulations, 
respectively.
We use small $h$ and large $\alpha$ to obtain large Schmidt numbers required 
for fluid-like behavior\cite{ripoll2004,ripoll2005}.
Most of the results shown are obtained from two-dimensional systems,
except in a few cases where this is explicitly mentioned.

In Tab.~\ref{tab:omegaH}, the theoretical values for the diffusion 
constant $D$ are given for $h=0.1$ for the different collision methods 
and various monomer densities \cite{ripoll2005,gg:gomp07xxe}.  
The corresponding results for other time steps $h$ can be obtained 
by employing the linear relationship between $D$ and $h$. 

\begin{table}[h]
  \begin{tabular}[t]{c|c|c|c}
    $\rho$ \ & $D^{\rm (SRD)}$ & $D^{{\rm (AT}-a{\rm)}}$ & \ $D^{{\rm (AT}+a{\rm)}}$ \\
    \hline
    10 \ \ & \ \ 0.1222 \ \ & \ \ 0.1222 \ \ & \ \ 0.1353\\
    20 \ \ & \ \ 0.1105 \ \ & \ \ 0.1105 \ \ & \ \ 0.1162\\
    40 \ \ & \ \ 0.1051 \ \ & \ \ 0.1051 \ \ & \ \ 0.1078\\
  \end{tabular}
  \caption{Diffusion constants $D$ of point-particle fluids
   for the standard MPC-SRD algorithm, as well as for 
   MPC-AT$-a$ and MPC-AT$+a$ simulations for various monomer densities,
   in two dimensions. All data are calculated for collision time $h=0.1$.
   Diffusion constants for other time steps $h$ can be obtained
   by employing the linear relationship between $D$ and $h$.   
   Note that the values for MPC-AT$-a$ are identical 
   with those for MPC-SRD with collision angle $\alpha=90^\mathrm{o}$.
}
  \label{tab:omegaH}
\end{table}

\subsection{Steady Shear Flow}
\label{sec:steadyflow}

\begin{figure}[ht]
  \begin{center}
    \includegraphics[width=7cm,clip]{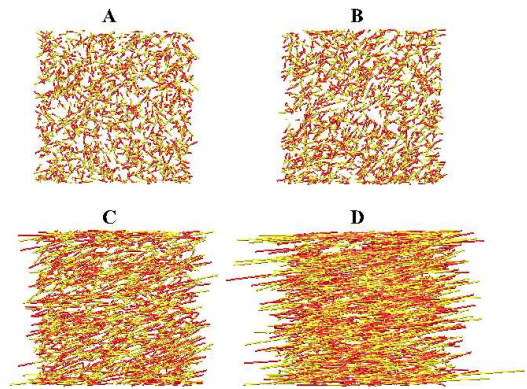}
  \end{center}
  \caption{
    \label{snap}
     (Color online) 
     Snapshots of dumbbell configurations in steady shear flow.
     The system size is $L_x=L_y=50$.
     Half of each dumbbell is colored red, the other half yellow for reason 
     of visualization.  In each frame only 2500 dumbbells are shown, 
     so that the density is $10$ times as high as appears from the pictures.
     The spring constant and the collision time are $\mathrm{K}=0.2$ and 
     $h=0.02$, respectively.
     From (A) to (D), the applied shear rates are 
     $\dg / \omega_H=0.0565$, $0.565$, $2.83$ and $5.65$.
       }
\end{figure}

In \Fig{snap}, we present snapshots for steady shear flow 
with a simulation box containing 25000 dumbbells.
At lower shear rates, \ie $\dg / \omega_H \leq 0.6$, 
see \Fig{snap}A and \ref{snap}B, the average extension of the dumbbells 
is hardly distinguishable from the equilibrium value. 
In these two cases, the shear flow is not strong enough 
to align the dumbbells along the flow direction, 
so that both systems are still isotropic.
With increasing $\dg$, shear forces overwhelm entropic forces.
As a result, dumbbells are largely stretched, 
at the same time reorientated along the flow direction, 
as presented in \Fig{snap}C and \ref{snap}D.
Note that near both the walls, the average size 
$\langle \mathrm{r^2} \rangle ^{1/2}$ 
of the dumbbells in flow is larger than in the bulk.
Also, an alignment of the dumbbells is found near 
the walls, both with and without shear flow, with peaks at $y=0$ and $y=L_y$.
This is an effect of the geometrical constraints imposed on anisotropic
particles by a hard wall.
Furthermore, a maximum of the extension occurs at a {\em finite} distance 
from the wall, which we attribute to the combined effect of the wall 
and the flow conditions; dumbbells very close to the wall
are sterically oriented parallel to the wall and thus experience only
a very small shear force, while those a little further away are close to
the average inclination angle (see \Fig{e2e} below), which corresponds 
to the largest stretching. The distance of the position of the maximum 
from the wall decreases with increasing shear rate, and seems to 
approach the size of the collision cells for large 
$\dot\gamma$. 
The relative peak height increases with increasing shear rate. 
For example, we find that the maximum extension
$\langle \mathrm{r^2} \rangle ^{1/2}$ near the wall is about $11\%$ larger 
than the bulk extension for $\dg/\omega_H=1.13$, while it is about 
$28\%$ larger than in the bulk for $\dg/\omega_H=2.83$. 

\begin{figure}[ht]
    \label{e2e}
  \begin{center}
    \includegraphics*[width=7cm,clip]{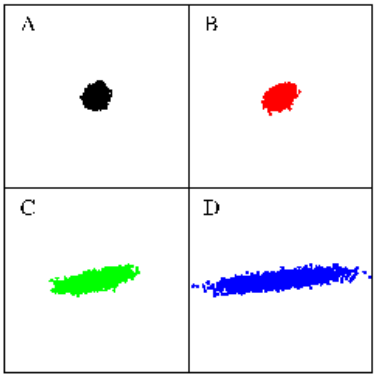}\\
  \end{center}
  \caption{
     (Color online) 
     Distribution of dumbbell configurations for the system shown 
     in Fig.~\ref{snap}. Each dot indicates the end-to-end vector 
     of a dumbbell. 
       }
\end{figure}

\begin{figure}[ht]
  \begin{center}
    \includegraphics*[clip]{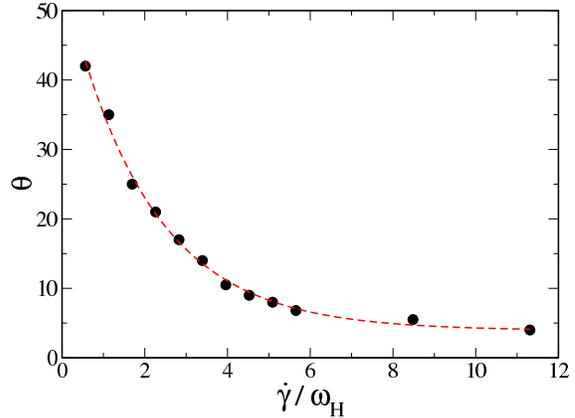}
  \end{center}
  \caption{
    \label{angle}
     (Color online) 
     The inclination angle $\theta$ as a function of dimensionless shear rate 
     $\dg / \omega_H$ for the system of \Fig{snap}, with spring constant 
     $\mathrm{K}=0.2$, collision time $h=0.02$, and system size $L_x=L_y=50$.
       }
\end{figure}

\Fig{e2e} presents the extensional and orientational distribution of the 
dumbbells for various shear rates. 
At lower shear rates, $\dg / \omega_H \leq 1$, the end-to-end vector 
of the dumbbells is distributed on a circle, see \Fig{e2e}A and B, 
indicating an isotropic orientation. 
At a higher shear rate, $\dg / \omega_H = 2.83$, 
the orientational distribution becomes an elongated ellipse, see \Fig{e2e}C.
With increasing shear rate, the distribution elongates further.
Simultaneously, dumbbells become more aligned with the flow direction, 
as can be seen quantitatively from the inclination angle $\theta$ shown in 
\Fig{angle}. Here, the inclination angle is defined as the angle between
the average orientation of the end-to-end vector of a dumbbell and the
the flow direction. 
At lower shear rates, $\dg / \omega_H \leq 1$, the inclination angle 
approaches $\theta=45^\mathrm{o}$,
while it decays to zero for large shear rates with a 
power law $\dot\gamma^{-1}$. 

\begin{figure}[ht]
  \begin{center}
    \includegraphics*[clip]{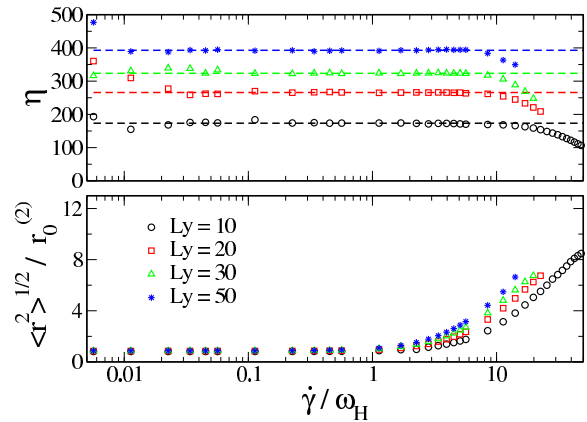}
  \end{center}
  \caption{
    \label{eta-r-gamma-y}
     (Color online) 
     Shear viscosity $\eta$ and scaled average dumbbell length  
     $\langle \mathrm{r}^2 \rangle^{1/2} / \mathrm{r}_0^{(2)}$ 
     as a function of dimensionless shear rate $\dg / \omega_H$.
     Systems with the wall separation 
     $L_y=10,~20,~30$, and $50$ are investigated.
     The spring constant is $\mathrm{K}=0.2$ and the collision time $h=0.02$.
       }
\end{figure}

In \Fig{eta-r-gamma-y}, we plot the shear viscosity $\eta$
as a function of dimensionless shear rate $\dg / \omega_H$ for various 
wall separations $L_y$ ranging from $10$ to $50$.
In each system, $\eta$ remains constant until the applied shear rate 
reaches a critical value $\dg_c / \omega_H \approx 5$. 
The shear viscosity then decays rapidly as $\dg$ further increases, 
showing a typical ``shear-thinning'' behavior.
\Fig{eta-r-gamma-y} also shows the average extension of dumbbells 
$\langle \mathrm{r}^2 \rangle^{1/2} / \mathrm{r}_0 ^{(2)}$ 
as a function of the shear rate.

Two comments are required here.  Firstly, in our MPC model, an entanglement 
between dumbbells is not taken into account, so that they can 
freely cross each other.
Also, the absence of an excluded-volume interaction implies that there
is no benefit of a para-nematic ordering in terms of an increased
sliding of parallel dumbbells along each other as in solutions of
rod-like colloids; instead, 
parallel dumbbells interact very similarly to isotropically 
oriented dumbbells, since in both cases the monomers colloide
with other monomers in exactly the same fashion. Thus, our system 
is very similar to a solution of non-interacting harmonic dumbbells,
for which -- in the absence of a finite extensibility -- 
neither ``shear-thinning" nor ``shear-thickening"
is expected\cite{bird87,doi86}, compare Eq.~(\ref{visc_theo0}). 
Secondly, the size of the simulation box should have no influence 
on the bulk viscosity at a given shear rate. 
However, the plateau value of the viscosity 
increases strongly with the wall separation $L_y$. 
This indicates that boundary effects could be responsible for the observed 
``shear-thinning'' behavior.

\begin{figure}[ht]
  \begin{center}
    \includegraphics[clip]{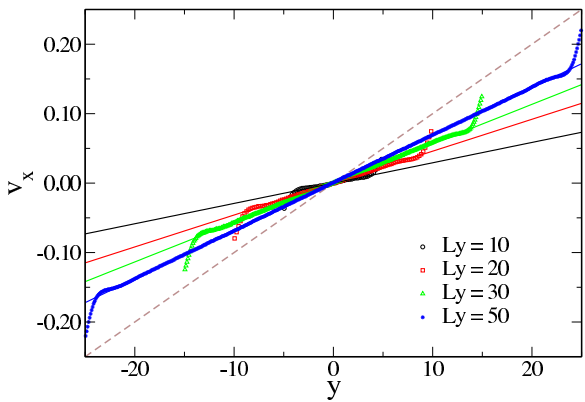}
  \end{center}
  \caption{
    \label{vp-y}
     (Color online) 
     Velocity profiles for wall separations 
     $L_y=10$, $20$, $30$, and $50$. Data are obtained for
     spring constant $\mathrm{K}=0.2$, collision time $h=0.02$, and shear 
     rate  $\dg / \omega_H=0.565$.
     The dashed line corresponds to the applied shear rate.
     Solid lines represent fits to the bulk part of the velocity profiles,
     their slopes yield the effective shear rates $\dg_{\mathrm{eff}}$.
       }
\end{figure}

\begin{figure}[ht]
  \begin{center}
    \includegraphics[clip]{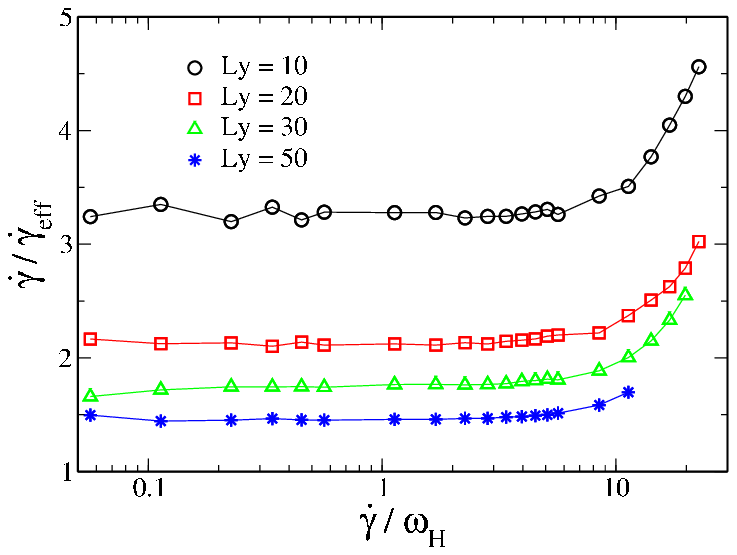}
  \end{center}
  \caption{
    \label{gamma-y}
     (Color online) 
     Ratios between the applied shear rates $\dg$ and the effective 
     shear rates $\dg_{\mathrm{eff}}$ as a function of $\dg/\omega_H$ 
     for the same systems as in \Fig{eta-r-gamma-y}.
  }
\end{figure}

We therefore examine the velocity profiles for systems with various wall 
separations $L_y$. In \Fig{vp-y}, the average velocities $v_x$ of the 
monomers along the 
flow direction  are plotted as function of $L_y$ for a fixed shear rate 
of $\dg / \omega_H = 0.565$. The velocities at the boundaries deviate 
only very little  from the wall velocities, \ie there is very little 
slip at the walls, as expected.
However, the velocity decays rapidly in a boundary layer of thickness $\Delta$,
then decays linearly to zero at the middle plane.
Obviously the applied shear rate $\dg$ is not appropriate to calculate 
the shear viscosities
from the stress tensor $\sigma_{xy}$ by $\eta = \sigma_{xy} / \dg$. 
An effective shear rate $\dg_{\mathrm{eff}}$ is therefore introduced instead, 
which characterizes the linear bulk part of the velocity profile.
At a given shear rate, the larger the wall separation,
the less the effective shear rate deviates from $\dg$,
since the finite-size effect is much stronger in smaller systems.
The ratio $\dg / \dg_{\mathrm{eff}}$ between the applied 
and the effective shear rates is plotted in \Fig{gamma-y},
as a function of $\dg /\omega_H$. 
At lower shear rates, \ie $\dg < \dg_\cc$, 
where $\dg_\cc$ is the critical shear rate,
the ratio $\dg / \dg_{\mathrm{eff}}$ is independent of the shear rate.
When the applied shear rate becomes larger than this critical value, 
the effective shear rate $\dg_{\mathrm{eff}}$ increases more slowly 
than $\dg$.

\begin{figure}[ht]
  \begin{center}
    \includegraphics[clip]{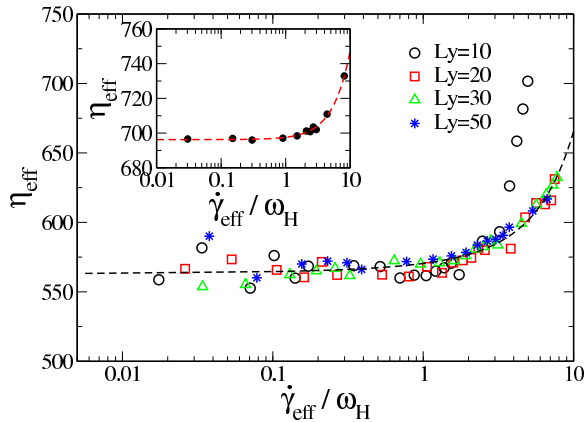}
  \end{center}
  \caption{
    \label{eta-eff-y}
     (Color online) 
     Master curve of the viscosity $\eta_\mathrm{eff}$ as a function of the
     effective shear rate $\dg_\mathrm{eff}$ on a semi-logarithmic scale. 
     Symbols are the same as in \Fig{eta-r-gamma-y} and \Fig{gamma-y}.
     For strong shear flow, \ie $\dg_\mathrm{eff} / \omega_H \gtrsim 2$, 
     the viscosity increases (``shear thickening'').
     The dashed line is fitted to the data with wall separations $L_y \geq 20$.
     In the inset, $\eta_\mathrm{eff}$ is plotted as a function of 
     $\dg_\mathrm{eff} / \omega_H$ for three-dimensional systems.
     A $20 \times 20 \times 10$ simulation box is chosen, 
     while the spring constant $\kk$, the collision time $h$ 
     and the average number density $\rho$ are 
     the same as in the two-dimensional systems.
  }
\end{figure}

Consequently, the effective shear viscosity can be calculated by 
\begin{equation}
\label{etaeff}
\eta_{\mathrm{eff}}=\frac{\sigma_{xy}}{\dg_\mathrm{eff}} \;.
\end{equation}
In \Fig{eta-eff-y}, $\eta_{\mathrm{eff}}$ is shown against 
$\dg_{\mathrm{eff}} / \omega_H$ for various wall separations $L_y$.
The data for different system sizes now all fall onto a single
master curve, which describes the bulk shear viscosity. 
Now, instead of ``shear thinning'' shown in \Fig{eta-r-gamma-y}, a 
very weak ``shear thickening'' behavior is observed when 
$\dg_{\mathrm{eff}} / \omega_H > 1$. 
Three-dimensional simulations are also carried out for systems of 
$20 \times 20 \times 10$ boxes along the $x$, $y$, and $z$ directions. 
For the same parameters $\mathrm{K}=0.2$ and $h=0.02$, 
weak ``shear thickening'' behavior is also observed, as shown in the inset 
of \Fig{eta-eff-y}, when $\dg_\mathrm{eff} / \omega_H$ reaches the critical 
value, $\dg_{\cc,\mathrm{eff}} / \omega_H \approx 2$. 
The value of the critical 
shear rates are found to be very similar in two and three dimensions.

\Fig{eta-eff-y} shows that the effective shear viscosity 
$\eta_\mathrm{eff}$ is nearly independent of the shear rate for 
$\dg_\mathrm{eff} /\omega_H \leq \dg_\mathrm{c,eff} /\omega_H \approx 2$.
This critical shear rate corresponds to the onset of the apparent 
``shear thinning'' observed in \Fig{eta-r-gamma-y}, as well as the 
deviation of 
$\dg / \dg_\mathrm{eff}$ from its low-shear-rate value in \Fig{gamma-y}.
It should be noticed that the value of 
$\dg_{\cc,\mathrm{eff}} / \omega_H \approx 2$
implies $\dg_{\cc} / \omega_H$ is in the range $[3,6.4]$ for 
system sizes $L_y \in [10,50]$, compare \Fig{gamma-y}.
However, it is important to note that there is already a pronounced 
alignment and stretching of the dumbbells for smaller shear rates;
\Fig{angle} shows that the inclination angle $\theta$ has decreased from 
$\theta=45^\mathrm{o}$ in the absence of shear flow to 
$\theta \approx 15^\mathrm{o}$ at $\dg /\omega_H=3$, while 
\Fig{eta-r-gamma-y} indicates that 
$\langle \mathrm{r}^2 \rangle^{1/2} / \mathrm{r}_0 ^{(2)} \approx 2 $ at 
$\dg /\omega_H=3$. 

\begin{figure}[ht]
  \begin{center}
    \includegraphics[clip]{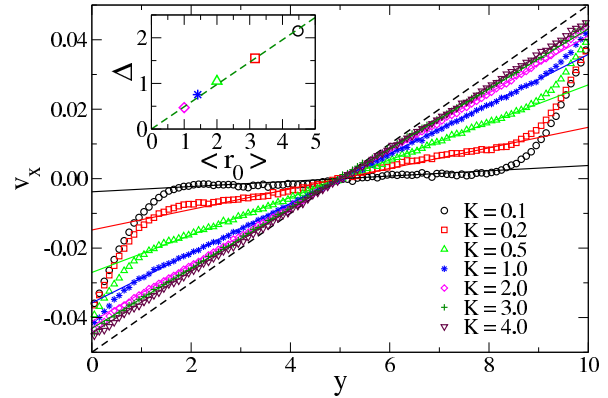}
  \end{center}
  \caption{
    \label{vp-k}
     (Color online) 
     Velocity profiles for various spring constants,  
     ranging from $\mathrm{K}=0.1$ to $\mathrm{K}=4.0$.
     The wall separation in each case is $L_y=10$. 
     The dashed line corresponds to the applied shear rate $\dg=0.01$, 
     while solid lines  are the fitted effective velocity profiles.
     The inset shows the thickness of the boundary layer $\Delta$ as a 
     function of equilibrium extension 
     $\mathrm{r}_0^{(2)}  = \sqrt{2\kt/\kk}$. The dashed line is a 
     linear fit.
       }
\end{figure}

The spring constant $\mathrm{K}$ of the dumbbells is of great importance,
since it controls the elasticity of the fluid.  We have therefore 
examined velocity profiles of systems of dumbbells with various 
spring constants.  In \Fig{vp-k}, the simulation results are plotted for 
a fixed applied shear rate $\dg=0.01$.
The effect of the boundary layer becomes more pronounced with decreasing 
spring constant.  By fitting the linear parts of the velocity profiles, 
we find that, for the same shear rate $\dg=0.01$,
the effective shear rate $\dg_\mathrm{eff}$ for dumbbells with $\mathrm{K}=0.1$
is about 10 times lower than that with the highest spring constant studied 
here, $\mathrm{K}=4.0$.  The thickness of the boundary layer is 
proportional to the equilibrium average extension 
$\mathrm{r}_0^{(2)} = \sqrt{2\kt/\kk}$, as shown in the inset of \Fig{vp-k}.

\begin{figure}[ht]
  \begin{center}
    \includegraphics[clip]{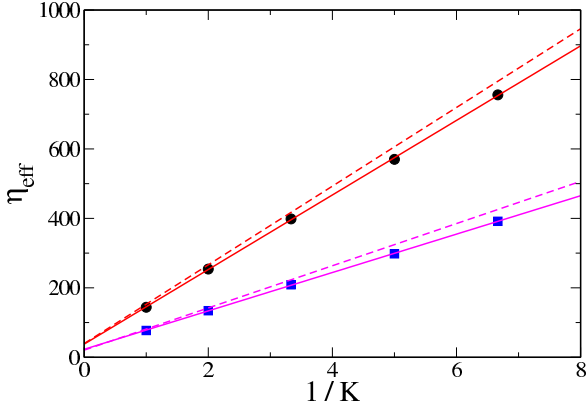}
  \end{center}
  \caption{
    \label{3d-k}
     (Color online) 
     The zero-shear viscosity $\eta_\mathrm{eff}$ 
     as a function of spring constant $\kk$  
     in both two-dimensional (circles) and three-dimensional 
     (squares) systems.
     The solid lines are linear fits, the dashed lines indicate
     the theoretical predictions (\ref{eq:visc_theo_MPC}). 
     In all simulations, the collision time is $h=0.02$.
     Two-dimensional simulations are performed in systems of 
     $50 \times 10$ boxes, while three-dimensional simulations are in 
     $30 \times 30 \times 20$ boxes
     along the $x$, $y$ and $z$ directions, respectively. 
     The average number density in three-dimensional systems is $\rho=10$, 
     which is half of value in two-dimensional systems.
           }
\end{figure}

The zero-shear viscosity $\eta_\mathrm{eff}$ is found to depend linearly 
on $1/\kk$, see \Fig{3d-k}.  As $\mathrm{K}$ increases,
the effective viscosity $\eta_\mathrm{eff}$ approaches the expected 
value of system of point particles with mass of $\mm^{\mathrm{c}}$ and
density $\rho/2$.  
The same linear relationship between $\eta_\mathrm{eff}$ and $1/\kk$ is 
also obtained in three-dimensional systems, as shown in \Fig{3d-k}.
Not only the linear dependence of $\eta_\mathrm{eff}$ on $1/\kk$ but
also the prefactors are in very good agreement with the theoretical
predictions (\ref{eq:visc_theo_MPC}).

\begin{figure}[ht]
  \begin{center}
    \includegraphics[clip]{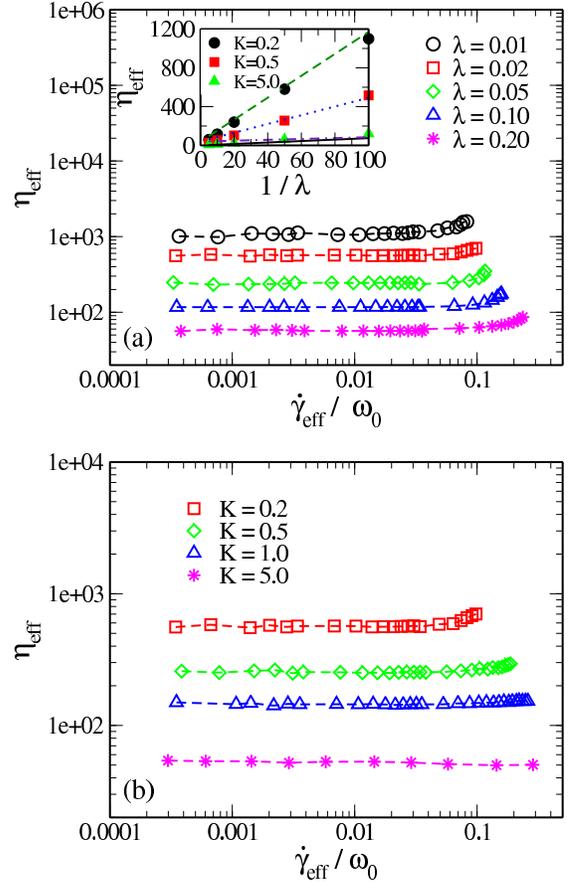}
  \end{center}
  \caption{
    \label{dt-y}
     (Color online) 
     The effective shear viscosity $\eta_\mathrm{eff}$ as a function of 
     the dimensionless effective shear rate $\dg_\mathrm{eff} / \omega_0$,
     on a double-logarithmic scale.
     (a) For fixed spring constant $\kk=0.2$ and various mean free paths 
     $\lambda=0.01$, $0.02$, $0.05$, $0.1$ and $0.2$.
     (b) For fixed mean free path $\lambda=0.02$ and various spring constants 
     $\kk=0.2$, $0.5$, $1.0$ and $5.0$.      
     In both cases, the wall separation is $L_y=10$.
     The inset in (a) shows the zero-shear viscosity $\eta_\mathrm{eff}$ 
     as a function of $1/\lambda$.
     The solid line indicates the theoretical result for 
     $\kk\rightarrow\infty$, while the other lines show the predictions 
     (\ref{eq:visc_theo_MPC}) for $\kk=0.2$, $0.5$, and $5.0$.
        }
\end{figure}

The scaled mean free path $\lambda$, 
which determines how far a point particle travels between collisions, 
is another important parameter which affects the shear viscosity.
We always employ small mean free paths\cite{ripoll2004,ripoll2005}, 
so that the collisional viscosity is dominant compared to the kinetic 
viscosity.  The data of \Fig{dt-y}(a) demonstrate that the zero-shear 
viscosity increases linearly with $1/\lambda$,
for all spring constants $\kk$ studied here, 
as it does for a system of point particles\cite{kikuchi2003,tuez03,ripoll2005}.
However, the slope decreases with increasing $\kk$, in good agreement 
with the analytical results obtained from Eq.~(\ref{eq:visc_theo_MPC}), 
as shown in the inset of \Fig{dt-y}(a).

The weak ``shear-thickening'' behavior is observed for all mean free 
paths investigated here, see \Fig{dt-y}. Thus, this weak 
``shear-thickening'' behavior is intrinsic to the MPC algorithm, 
and cannot be avoided by a variation of the collision time.
\Fig{dt-y}(a) indicates that the critical shear rate 
$\dg_\mathrm{c,eff} /\omega_0$ depends only very weakly on the
mean free path $\lambda$. Therefore, we present the simulation
data in \Fig{dt-y} as a function of $\dg_\mathrm{eff} /\omega_0$,
since $\omega_0=(2\kk/m)^{1/2}$ is independent of $\lambda$, while 
$\omega_H$ decreases linearly with $\lambda$. 
The ``shear-thickening'' behavior becomes more pronounced and 
slowly shifts to smaller values of $\dg_\mathrm{eff} /\omega_0$ 
for system of dumbbells with smaller spring constants, see \Fig{dt-y}(b). 
In the range of investigated spring constants and mean free paths,
the shear thickening occurs roughly at 
$\dg_\mathrm{c,eff} /\omega_0 \simeq 0.1$.
It is important to note that
the viscosity of the standard point-particle MPC fluid is also not
independent of the shear rate, but shows a weak {\em shear-thinning} behavior
at high shear rates \cite{kikuchi2003}. For our model parameters and
in two dimensions, this shear-thinning behavior sets in at a shear 
rate $\dg_\mathrm{c} \simeq 1$. Thus, with increasing $\kk$, shear
thickening occurs at a slowly increasing 
$\dg_\mathrm{c,eff} /\omega_0$ for $\kk\le 1$; for larger spring
constants $\kk \ge 5$, shear thinning is observed instead, and
$\dg_\mathrm{c,eff} /\omega_0$ decreases again (since 
$\dg_\mathrm{c,eff}\to 1$ and $\omega_0 \to \infty$ for $\kk\to \infty$).

\subsection{Small-amplitude Oscillatory Shear Flow}

Another way to explore the viscoelastic properties of a fluid is
to apply a small-amplitude oscillatory shear flow. 
We use here the strain amplitudes $\gamma_0 = \dg / \omega$ in the range $0.1$ to $0.5$ to mimic a 
small amplitude shearing. 
The frequencies $ \omega$ ranges from $10^{-4}$ to $10^{-1}$ in our simulations, which 
provides a wide range of shear rate from $10^{-5}$ to $5 \times 10^{-2}$.

\begin{figure}[ht]
  \begin{center}
    \includegraphics[clip]{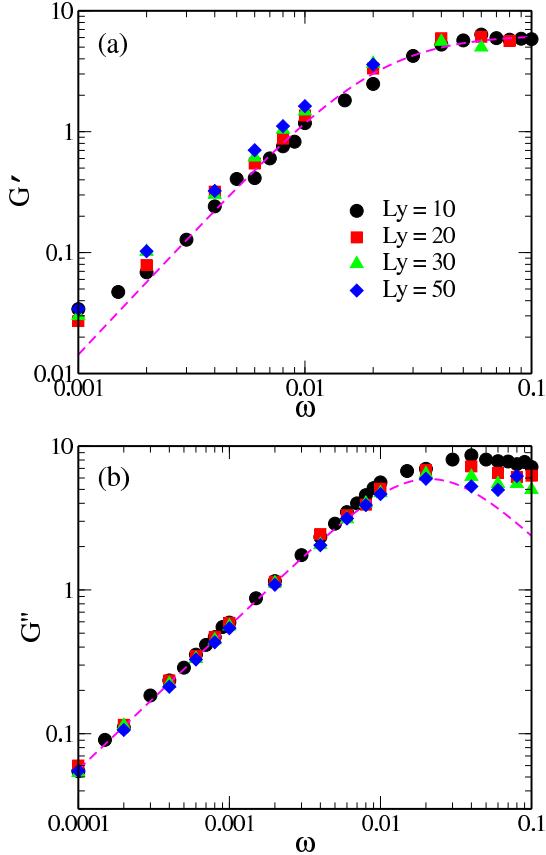}
  \end{center}
  \caption{
    \label{y-g}
     (Color online) 
     (a) Storage $G'$ and (b) loss modulus $G''$, 
     as a function of oscillation 
     frequency $\omega$ on a double-logarithmic scale, for systems with 
     various wall separations ranging from $L_y=10$ to $50$. 
     The spring constant and the collision time are $\mathrm{K}=0.2$ and 
     $h=0.02$, respectively.
     The dashed line in (a) is fitted by the Maxwell model, \Eq{mg1}, on 
     the basis of all simulation data,
     while the one shown in (b) is based on data
     for oscillation frequencies $\omega < 0.02$.
         }
\end{figure}

The storage and loss moduli as a function of oscillation frequency are plotted in \Fig{y-g}.
Similarly to the simulations of steady shear flow, 
effective shear rates are measured from 
the bulk velocity profiles at times when $\cos(\omega t)=\pm1$.
By doing so, all the simulation data fall onto master curves 
at various wall separations from $L_y=10$ to $50$. 
As can been seen from \Fig{y-g}(a), the storage modulus $G'$ is well 
fitted by \Eq{mg1}, indicating that the dumbbell system exhibits a 
typical behavior of a Maxwell fluid.
The relaxation frequency $\omega^*$ obtained from the fit of the storage 
modulus $G'$ against $\omega$  is then used in \Eq{mg2} to fit
the loss modulus $G''$.
In \Fig{y-g}(b), at low frequencies, $\omega \leq 0.02$,   
the simulation data follow the expected linear $\omega$-dependence very well.
In this linear regime, the shear viscosity is then calculated by 
$\eta = G''(\omega) / \omega$, which yields $\eta = 565$, 
in excellent agreement with the result in steady shear flow, 
compare \Fig{eta-eff-y}.
Note that the fitted values for the amplitude $G^*$ in \Eqs{mg1} and \eq{mg2} 
differ by about a factor 2.
This indicates that the system investigated here does not behave exactly like a simple Maxwell fluid.

\begin{figure}[ht]
  \begin{center}
    \includegraphics[clip]{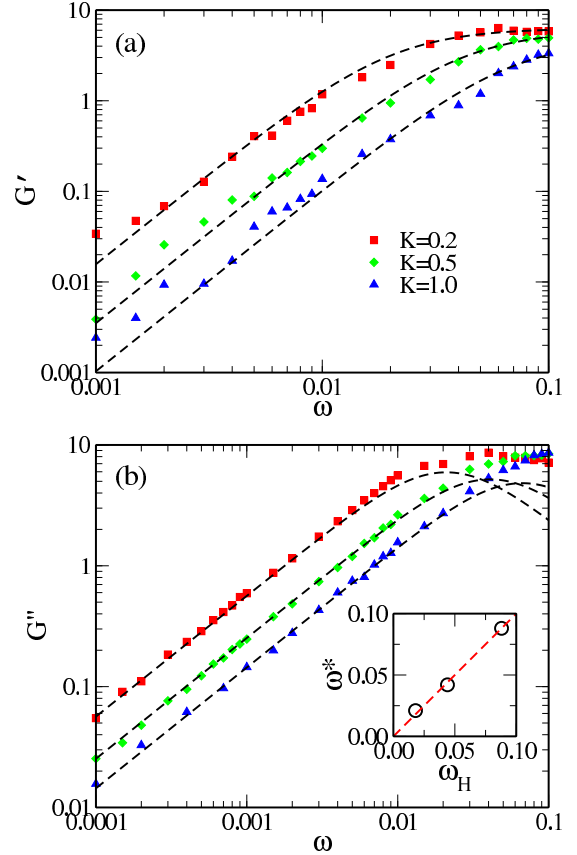}
  \end{center}
  \caption{
    \label{k-g}
     (Color online) 
     (a) The storage $G'$ and (b) the loss moduli $G''$,
     as function of oscillation 
     frequency $\omega$ on a double-logarithmic scale, for systems of 
     dumbbells with various spring constants ranging from 
     $\mathrm{K}=0.2$ to $\mathrm{K}=1.0$.
     The wall separation and the collision time are $L_y=10$ and $h=0.02$, 
     respectively. The inset shows the fitted relaxation frequencies 
     $\omega^*$ as a function of the frequency $\omega_H$ predicted by 
     Eq.~(\ref{eq:omega_H}). The dashed line shows the identity 
     $\omega^*=\omega_H$. 
        }
\end{figure}

In \Fig{k-g}, we examine the storage and loss moduli of system of dumbbells 
with various spring constants.
As in \Fig{y-g}, simulation results are all well fitted by 
the Maxwell equations \eq{mg1} and \eq{mg2}, except for somewhat different 
amplitudes $G^*$. The relaxation frequency $\omega^*$ is found 
to agree very well with $\omega_H$, as shown in the inset of \Fig{k-g}.
At lower oscillation frequency in \Fig{k-g}(b), the viscosities calculated 
from $G''(\omega)/ \omega$ are $\eta=565$, $253$ and $144$
for systems with $\kk=0.2$, $0.5$ and $1.0$, respectively.
These values are again in excellent agreement with those calculated from 
\Eq{etaeff} in steady shear flow.  For all spring constants $\kk$, 
the fitted amplitudes $G^*$ for the storage moduli $G'$ are about half of
those calculated for the loss moduli $G''$.
This indicates that even for a system of dumbbell with high spring constant, 
a simple Maxwell model is not appropriate for a quantitative description.


\subsection{Angular Momentum Conservation}

\begin{figure}[ht]
  \begin{center}
    \includegraphics[clip]{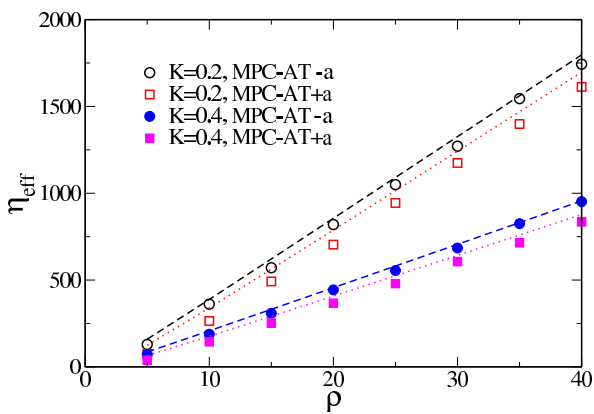}
  \end{center}
  \caption{
    \label{fig:eta_rho}
     (Color online) 
     Effective shear viscosities $\eta_{\rm eff}$ as a function of the 
     density $\rho$ for MPC-AT$-a$ and MPC-AT$+a$, each for 
     spring constants $\kk=0.2$ and $\kk=0.4$.
     The lines represent the theoretical results obtained from 
     Eq.~(\ref{eq:visc_theo_MPC}).
     The wall separation and the collision time are $L_y=20$ and 
     $h=0.014$, respectively.  
       }
\end{figure}

The viscosity of a simple MPC-AT$+a$ fluid (with angular-momentum
conservation) is about a factor 
$1/2$ smaller than of a MPC-AT$-a$ fluid \cite{gg:gomp07h,gg:gomp07xxe}.
We thus expect the viscosity of the dumbbell fluid to be affected 
by angular-momentum conservation as well.
The simulation results for both MPC-AT$-a$ and MPC-AT$+a$ methods are 
compared in Fig.~\ref{fig:eta_rho}. We find that the effective zero-shear 
viscosity $\eta_{\rm eff}$ increases linearly with the monomer density 
$\rho$ for $\rho \gtrsim 5$.
The corresponding theoretical results (\ref{eq:visc_theo_MPC}) are in 
good agreement with the simulation results for both investigated spring
constants.
Minor deviations from the linear relationship of $\eta_{\rm eff}$ with 
$\rho$ originate from the variation of the diffusion constant at low 
densities, which approaches a constant value for high $\rho$.
The viscosity of MPC-AT$+a$ is lower than for MPC-AT$-a$, although 
this effect is less pronounced than for pure point-particle MPC fluids, 
since the main contribution to the viscosity originates from the spring
tension.

\begin{figure}[ht]
  \begin{center}
    \includegraphics[clip]{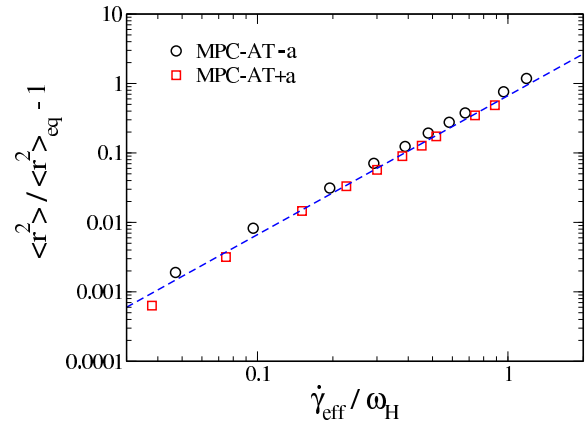}
  \end{center}
  \caption{
    \label{fig:R2}
     (Color online) 
     Scaled average of the dumbbell extension, 
     $\langle r^2 \rangle / \langle r^2 \rangle_{\rm eq} -1$,
     as a function of the effective 
     shear rate $\dg_\mathrm{eff}/\omega_H$ for angular-momentum conserving 
     and non-conserving methods.
     The spring constant and collision time are $\kk=0.2$ and 
     time step $h=0.014$, respectively, the density is $\rho=10$ and
     the wall separation is $L_y=20$.
     The dashed line represents the theoretical result (\ref{eq:R2}).
       }
\end{figure}

In Fig.~\ref{fig:R2}, we present the average squared dumbbell extension,
determined in the bulk as a function of the effective shear rate, along 
with the theoretical results (\ref{eq:R2}), for both the MPC-AT$-a$ and 
MPC-AT$+a$ methods. 
Note that the diffusion constant $D$ in Eq.~(\ref{eq:R2}) is different 
for angular-momentum conserving and non-conserving methods.
The angular-momentum conservation has only little effect on the spring 
extension; for a given effective shear rate, the extension is slightly 
lower for the angular-momentum conserving method.
The agreement of the simulation data with the theoretical
result (\ref{eq:R2}) is again remarkably good.


\section{Discussion}

As a further test for the correct calculation of the effective viscosity
by the procedure described in Secs.~\ref{sec:stress} and 
\ref{sec:steadyflow}, we have also determined the 
viscosity from Poiseuille flow. As in Ref.~\onlinecite{gg:gomp02c}, we apply
a gravitational force of strength $g$ parallel to the walls, with $g$ in 
the range from $g=0.0001$ to $g=0.01$ (in units of $\kt/a_0$). 
We fit the central part of the
velocity profile to a parabolic flow curve. The value of this curve at
the wall positions determines the effective wall slip. When this
slip velocity is subtracted, Eq.~(\ref{Poiseuille}) in 
Sec.~\ref{sec:stress} is employed to determine the viscosity\cite{zhang2007}.
We have used this method for a system of dumbbells with $\kk=0.2$ 
in a $30 \times 30$ box.  Excellent agreement between the two methods 
to calculate the zero-shear viscosity is obtained.  

Our results for the dependence of the inclination angle $\theta$ on the 
shear rate can be compared with the decay
of the inclination angle of flexible and semi-flexible
polymers. For dilute polymer solutions in the asymptotic regime of 
high shear rates (where the finite extensibility is important), $\theta$ 
has been predicted from Brownian dynamics simulations \cite{schr05} and 
theory \cite{wink06} to decay with a power law $\dot\gamma^{-0.3}$
and $\dot\gamma^{-1/3}$, respectively. For extensible dumbbells, the theory of 
Ref.~\onlinecite{wink06} predicts\cite{wink07} $\theta\sim\dot\gamma^{-1}$, in 
excellent agreement with our simulation results.

The wall slip in polymer melts has been studied extensively. In this case,
molecular dynamics simulations of polymer fluids with Lennard-Jones interactions
between monomers give a wall slip with a boundary layer thickness, which is on
the order of the monomer diameter $\sigma$ or less \cite{prie04,zhang2007}. Our 
model could be compared more easily with results for polymer solutions,
because our model does not include excluded-volume interactions. However,
there is little knowledge about semi-dilute polymer solutions near a wall
under flow conditions. Nevertheless, some comparisons with polymer melts with
moderate chain lengths are possible, where entanglement effects are absent.
For example, the molecular dynamics simulations of Zhang et al.\cite{zhang2007}
show a maximum of mean squared radius of gyration at a finite distance 
$\Delta_m$ from the wall, which shifts from $\Delta_m \simeq 1.5 \sigma$ 
for chains with $4$ monomers to $\Delta_m \simeq 2.2 \sigma$ for 
$10$ monomers. 


\section{Summary and Conclusions}

A multi-particle collision dynamics (MPC) algorithm has been developed 
to investigate the viscoelastic properties of harmonic-dumbbell fluid 
in shear flow.  The method is based on alternating streaming and 
collision steps, just as the original MPC method for Newtonian fluids. 
The only modification is to replace the ballistic motion of fluid
point particles by harmonic oscillations during the streaming step.
In this model, the entanglement between dumbbells is neglected.
Moreover, the storage and loss moduli are calculated 
by introducing a small amplitude oscillatory shear flow.

Our results can be summarized as follows:

First,
under steady shear flow, the dumbbells keep their isotropic distribution 
at low shear rates, but get highly stretched and orientated along the 
flow direction at high shear rates. 
The velocity profile is not uniform along the gradient direction, but
boundary layers of high shear develop near the walls. The thickness of
these boundary layers is found to scale with the size of the 
dumbbells in the absence of flow.  
The effective shear viscosity, calculated from the ratio 
between the off-diagonal component of the stress tenor, $\sigma_{xy}$,
and the effective shear rate $\dg_{\mathrm{eff}}$, 
expresses a very weak ``shear thickening'' behavior at high shear rates.

Second, 
the dependence of the viscosity on two parameters, the spring constant 
$\kk$ of the dumbbells and the collision time $h$, has been investigated.
These two parameters are of central importance, since the former 
controls the elastic energy of the system, while the latter determines 
the mean free path $\lambda$, which measures the fraction a the cell size
that a fluid particle travels on average between collisions.
We find that the shear viscosity of the dumbbell fluid increases 
linearly with $1/\kk$ and with $h$.

Third,
the storage and loss moduli of our viscoelastic solvent are studied 
by imposing an oscillatory velocity on the two solid walls.
The storage modulus $G'$ is found to be proportional to $\omega^2$ 
at low frequencies, and to level off at $\omega^*=\omega_H$. 
Its behavior over the whole frequency range studied here is well
described by a Maxwell fluid. 
The loss modulus $G''$ increases linearly with $\omega$ for low
frequencies.  The shear viscosities obtained from the ratio 
$G''/\omega$ at low shear rates agree very well with 
those obtained from simulations with steady shear. 
On the other hand, for $\omega > \omega_H$, we find that the data
approach a plateau value, while for a Maxwell fluid $G''$ would 
decrease again for higher frequencies. 

Our numerical results are quantitatively in good agreement with a 
simple theory, 
based on the kinetic theory of dilute solutions of dumbbells, where  
the transport coefficients of the standard MPC point-particle fluid 
are employed for the viscosity and the diffusion constant of the 
solvent. 

In our MPC algorithm of harmonic dumbbells, both elastic and viscous 
behaviors of solvent particles can be modeled properly, 
while hydrodynamic interactions are efficiently taken into account.
These are valuable assets to guide future simulations 
on investigating rheological properties of suspensions of 
spherical, rod-like or polymeric solute molecules in viscoelastic fluids.

\section*{Acknowledgment}

We thank R.G.~Winkler, M.~Ripoll and H.~Noguchi for 
many stimulating and helpful discussions.  Partial support of this work 
by the DFG through the Sonderforschungsbereich TR6 ``Physics of 
Colloidal Dispersion in External Fields'' is gratefully acknowledged.


\section*{Appendix: Analytical solution of the density profile
with attractive wall potentials\label{App}}

Combining Equations \eq{wall1} and \eq{wall2}, the density profile,
when attractive wall potentials are introduced, can be solved analytically.
Considering the symmetry of the density profile, $\rho(y)=\rho(L_y-y)$, 
only the initial half part need to be taken into account.
For $0 < y < 2c_1\mathrm{r}_0^{(1)}$,
we then arrive at
\begin{widetext}
\begin{eqnarray}
 \rho(y)  & = & \frac{1}{Z}  \left\{  \int^{L_y}_{0} d y'~
 \mathrm{e}^{- \kk (y-y')^2 / 2\kt} \right.
 \nonumber \\
 & + &  
 \left. \int^{2c_1\mathrm{r}_0^{(1)}-y}_{0} d y'~ 
 \mathrm{e}^{-\- \kk (y-y')^2 / 2\kt} ~
 \left[~\mathrm{e}^{ 2c_2 \kt (1 - (y+y')/2c_1\mathrm{r}_0^{(1)} ~)} -1~\right]  \right\} \;. 
 \nonumber  
\end{eqnarray}
which implies
\begin{eqnarray}
\rho(y) & = & \frac{1}{Z} \left\{ \mathrm{erf}\left(\sqrt{\kk /2\kt}~(L_y-y)\right)
         + \mathrm{erf}\left(\sqrt{\kk /2 \kt}~(2y-2c_1\mathrm{r}_0^{(1)})\right) \right.
 \nonumber \\
 &+& \left. \exp\left[\frac{(c_2\kt\mathrm{r}_0^{(1)})^2 + 
 4c_2\kt(c_1\mathrm{r}_0^{(1)}-y)}{2c_1\mathrm{r}_0^{(1)}}\right]
  \left[\mathrm{erf}\left( \frac{-c_2\kt + c_1\mathrm{r}_0^{(1)}y \kk /\kt}
  {c_1\mathrm{r}_0^{(1)} \sqrt{2\kk / \kt}} \right) \right. \right.
  \nonumber \\
 &+& \left. \left. \mathrm{erf}\left(\frac{ c_2\kt + 2(c_1\mathrm{r}_0^{(1)}-y) c_1\mathrm{r}_0^{(1)} \kk/ \kt}
  {c_1\mathrm{r}_0^{(1)} \sqrt{2\kk / \kt}} \right) \right]
  \right\} \;,
  \nonumber 
\end{eqnarray}
\end{widetext}
while for $2c_1\mathrm{r}_0^{(1)} < y < L_y/2$, the density profile is given by \Eq{wall0}.


\end{document}